\documentclass[twocolumn,amsmath,nofootinbib,amssymb]{revtex4}

\usepackage{graphicx}
\usepackage{dcolumn}
\usepackage{bm}

\newcommand{\lsim}{\mbox{\raisebox{-.9ex}{~$\stackrel{\mbox{$<$}}{\sim}$~}}}
\newcommand{\gsim}{\mbox{\raisebox{-.9ex}{~$\stackrel{\mbox{$>$}}{\sim}$~}}}
\newcommand{\calp}{{\cal P}}

\def\thebiblio#1{
\begin{center}\bf \large References
\end{center}
\list
{[\arabic{enumi}]}{\settowidth\labelwidth{#1.}\leftmargin\labelwidth
 \advance\leftmargin\labelsep
 \usecounter{enumi}}
 \def\newblock{\hskip .11em plus .33em minus -.07em}
 \sloppy
 \sfcode`\.=1000\relax}

\begin{document}

\title{Correlated curvature perturbations and magnetogenesis 
from the GUT gauge bosons}



\author
{\large Konstantinos~Dimopoulos}%
\email{k.dimopoulos1@lancaster.ac.uk}
\affiliation{Consortium for Fundamental Physics,
Physics Department, Lancaster University, Lancaster LA1 4YB, UK}%

\date{\today}

\begin{abstract}
In the context of hybrid inflation at the scale of grand unification,
it is shown that, if (some of) the gauge bosons of the unbroken grand unified 
group obtain a flat superhorizon spectrum of perturbations from inflation they 
can affect or even generate the observed curvature perturbation and 
simultaneously produce a sizable primordial magnetic field capable of 
explaining galactic magnetism. The mechanism employs the vector curvaton idea 
for a supermassive grand unified gauge boson after the breaking of grand 
unification.
\end{abstract}


\pacs{98.80.Cq}
\maketitle



Observations provide strong evidence that the Universe underwent a phase of
inflation in its early history. One of the most important consequences of 
inflation is the generation of the curvature perturbation $\zeta$, which is 
necessary for structure formation and is observed through the CMB anisotropy 
\cite{book}. Quantum fluctuations of suitable fields give rise to a flat 
superhorizon spectrum of perturbations through the process of particle 
production \cite{hawking}. Under certain circumstances these perturbations can 
create the curvature perturbation of the Universe. Until recently, only scalar 
fields have been employed for this task. In recent works, however, it has been 
shown that vector fields can also affect (by producing statistical anisotropy) 
or even generate $\zeta$ \cite{vecurv,yokosoda,stanis,fnlanis,anisinf,varkin}. 
The first such suggestion was made in the pioneering Ref.~\cite{vecurv}, where 
it was shown that, if an Abelian vector boson field obtains a flat superhorizon
spectrum of perturbations during inflation, it can act as a curvaton field 
\cite{curv} provided, after inflation, its mass-squared becomes positive and 
bigger than the Hubble scale. In this case, the vector field oscillates 
coherently, behaving as pressureless {\em isotropic} matter \cite{vecurv}. 
Thus, it can dominate the radiation background without introducing significant 
anisotropic stress, contributing thereby its own curvature perturbation 
according to the curvaton mechanism~\cite{curv}.
 
In this letter we examine the possibility that the supermassive bosons of a
Grand Unified Theory (GUT) can act as vector curvatons and contribute 
significantly to or even generate $\zeta$. We consider hybrid inflation, which 
ends at the breaking of grand unification, with the GUT Higgs playing the role 
of the waterfall field \cite{hybrid}. We also assume that, through some 
mechanism, the conformal invariance of (at least some of) the GUT gauge bosons 
is appropriately broken during inflation so that they obtain a nearly scale 
invariant spectrum of superhorizon perturbations. At the breaking of grand 
unification the supermassive gauge bosons, which couple to the GUT Higgs, 
become massive and undergo coherent oscillations before their decay. If they 
carry with them this perturbation spectrum, then they can play the role of 
vector curvatons.

As a side effect, the mechanism generates a primordial hypermagnetic field
due to the projection of the original perturbation spectrum onto the 
hypercharge direction. The hypermagnetic field is frozen into the
plasma of the reheated Universe and survives until much later times.
Eventually, it transforms into a regular primordial magnetic field at the 
electroweak phase transition, when it is projected onto the photon direction 
through the Weinberg angle.
In this letter we estimate the characteristics of this magnetic field and we 
find that, when the contribution to the curvature perturbation due to the 
supermassive GUT bosons is significant, the primordial magnetic field is strong
and coherent enough to account for the observed magnetic fields of the 
galaxies. Hence, through the vector curvaton mechanism we can connect directly 
structure formation and galactic magnetism.

Most of the observed galaxies carry $\mu$G magnetic fields, which are 
dynamically important \cite{kron}. In spirals the magnetic field follows the 
spiral arm structure, which indicates that it is being rearranged by a dynamo 
mechanism \cite{beck}. Such a galactic dynamo can amplify a weak seed field up 
to $\mu$G strength. 
The origin of this seed field, however, is still under debate. It has been 
theorised that the seed magnetic field can be of cosmological origin 
\cite{pmfrev}. 
The coherence required in order not to destabilise the galactic dynamo action
is attained if the magnetic field is produced during inflation through particle
production. This is possible, however, only if the conformality of 
electromagnetism is violated during inflation. Since the pioneering work in 
Ref.~\cite{TW}, a number of mechanisms have been suggested to this end 
\cite{mine,pmfinf,gaugekin}. 
Successful mechanisms manage to generate a superhorizon spectrum of photon 
perturbations giving rise to a superhorizon spectrum of magnetic fields.%
\footnote{Despite what is assumed in Ref.~\cite{mukh}, vector field 
backreaction does not spoil inflation but instead it can lead to a new 
inflationary attractor \cite{attr}, where steep inflation is allowed and the 
$\eta$-problem is overcome \cite{jack} (see also Ref.~\cite{sodarev} and 
references within).}

The above mechanisms can be employed in general to obtain a superhorizon 
spectrum of perturbations for Abelian gauge fields at the end of inflation. To 
have a significant contribution (for example produce statistical anisotropy) to
the curvature perturbation at observable scales such a spectrum needs to 
satisfy a further requirement; namely approximate scale-invariance. Without 
going into details for its origin, let us assume that, near the end of 
inflation, the spatial part of (some of) the unbroken GUT bosons $W_\mu$ has 
obtained a flat superhorizon spectrum of perturbations.

Our hybrid inflation ends through the GUT phase transition, in which some of 
the GUT gauge bosons become massive through their coupling to the GUT Higgs, 
with mass \mbox{$m_X=h M$}, where $h$ is the GUT coupling and 
\mbox{$M\sim 10^{16}\,$GeV} is the vacuum expectation value (VEV) of the Higgs 
field. These supermassive GUT bosons $X_\mu$ are produced from the mixing of 
some of the $W_\mu$ gauge bosons of the unbroken GUT, which carry the 
superhorizon perturbation spectrum. Thus, we would expect the $X_\mu$ to carry 
this spectrum of perturbations too. 
The supermassive $X$-bosons decay into lighter particles with decay rate 
\mbox{$\Gamma_X=\frac{h^2}{8\pi}m_X=\frac{h^3}{8\pi}M$}. This decay rate is 
comparable or smaller than the Hubble scale at the GUT transition
\mbox{$H_*\simeq\frac{1}{\sqrt 3} M^2/m_P$} if 
\mbox{$h\lsim(\frac{8\pi}{\sqrt 3}\frac{M}{m_P})^{1/3}\approx 0.5$},
where \mbox{$m_P=2.4\times 10^{18}\,$GeV} is the reduced Planck mass. In this 
case, the oscillating $X$-boson can affect the Universe expansion before its 
decay since its lifetime is comparable or larger than the timescale of the 
Universe dynamics. Hence, the $X$-boson can play the role of the vector 
curvaton.

The GUT bosons which are not coupled to the GUT Higgs remain massless and 
form the gluons and the electroweak gauge bosons. The latter also, carry the 
superhorizon spectrum of perturbations, which was generated during inflation.
The temperature of the plasma at the end of inflation is rather large 
\mbox{$T_{\rm end}\sim(m_P^2\Gamma H_*)^{1/4}$} \cite{KT}, where $\Gamma$ is 
the decay rate of the inflaton field. Typically, this temperature is much 
higher than that of the electroweak unification. The non-Abelian gauge fields, 
through their interaction with the hot plasma, obtain a thermal mass and become
screened \cite{nonAbelmass}. Thus, the superhorizon spectrum of their 
perturbations cannot survive. In contrast, the superhorizon perturbations of 
the Abelian Hypercharge field $Y_\mu$ are preserved.

Indeed, the conductivity of the plasma is very high, which means that the
hypermagnetic field \mbox{{\boldmath $B$}$^Y\equiv\,$
{\boldmath $\nabla$}$\,\times\,\delta${\boldmath $Y$}} 
becomes frozen into the plasma.
This guarantees that the memory of the superhorizon perturbations of the field 
is preserved. At the electroweak phase transition the hypercharge projects onto
the photon direction through the Weinberg angle $\theta_W$ as 
\mbox{{\boldmath $A$}$\,=\mbox{\boldmath $Y$}\cos\theta_W$}. This transforms 
the hypermagnetic field into a regular magnetic field which can become the seed
for galactic magnetism.

The equation of motion for the 
homogeneous zero-mode of the
supermassive $X$-boson in a spatially flat FRW Universe is
%
%
\begin{equation}
\ddot X+H\dot X+m_X^2 X=J\,,
\label{Xeom}
\end{equation}
where \mbox{$X\equiv|\mbox{\boldmath $X$}|$},
$J$ is a source current, and the dot denotes derivative with 
respect to the cosmic time $t$. 
The $X$-boson condensate, once formed at the GUT transition, 
begins coherent oscillations because its mass 
\mbox{$m_X\simeq hM\sim 10^{15}\,$GeV} is larger than 
the Hubble scale at the time 
\mbox{$H_*\sim\frac{1}{\sqrt 3}M^2/m_P\sim 10^{14}\,$GeV} so the
friction term in Eq.~(\ref{Xeom}) is subdominant. 

The $X$-boson decays quickly, since its decay rate is 
\mbox{$\Gamma_X\sim H_*$} (\mbox{$\Gamma_X\simeq H_*$ for $h\simeq 0.5$}). Before decaying, however, it manages to undergo 
several oscillations since \mbox{$m_X/\Gamma_X=\frac{8\pi}{h^2}\simeq 100$}. 
Note that, during these oscillations we 
can ignore the source current since the condensate evaporates no earlier than 
at the time of decay \mbox{$\Delta t_{\rm evap}\gsim\Gamma_X^{-1}$}. Before 
evaporation the effect of the source current is subdominant to the zero-mode, 
in the condensate's dynamics.

As shown in Ref.~\cite{vecurv}, a coherently oscillating vector field 
condensate behaves like pressureless, isotropic matter. This means that the 
oscillating $X$-boson can act as a curvaton if its density is near domination 
at the phase transition without producing significant anisotropic stress 
\cite{vecurv}. 

Defining the density parameter $\Omega_X$ at the end of inflation
(denoted by `end') as 
\mbox{$\Omega_X\equiv(\rho_X/\rho)_{\rm end}$}, we find
\begin{equation}
X_{\rm end}=\sqrt{6\Omega_X}\;\frac{m_PH_*}{m_X}\,,
\label{Xew}
\end{equation}
where the density of the Universe at the end of inflation is 
\mbox{$\rho_{\rm end}=3m_P^2H_*^2$} and we used that the density of the 
oscillating $X$-boson at that time 
is \mbox{$(\rho_X)_{\rm end}=\frac12 m_X^2 X_{\rm end}^2$}.
On energy equipartition grounds we expect \mbox{$\Omega_X\sim 0.1$}, so the 
condensate nearly dominates the Universe before decay, as required by the 
curvaton scenario.\footnote{%
The mass of the $X$-boson is determined by the GUT Higgs 
\mbox{$m_X(\phi)=h\phi$}. At the phase transition $\phi$ also oscillates 
briefly around its VEV. 
Consequently, Eq.~(\ref{Xeom}) can result in resonant amplification of the
amplitude of the condensate as \mbox{$X\sim e^{\mu m_\phi t}X_0$}, where 
\mbox{$X_0$} is the initial value and $\mu$ is the strength of the resonance 
(Floquet index). This ensures that $\rho_X$ rapidly 
becomes a significant fraction of the total density.}

The particle production process for vector fields is, in general, anisotropic 
because vector fields have several degrees of freedom which can undergo
particle production with different efficiency. If such fields contribute to the
curvature perturbation $\zeta$ then they may generate statistical anisotropy in
the spectrum $\calp_\zeta$ 
of the curvature perturbation \cite{stanis}. The latter is quantified by the 
so-called anisotropy parameter $g$.
As shown in Ref.~\cite{stanis}, 
\mbox{$g=p N_A^2\calp_+/\calp_\zeta^{\rm iso}$}, where $\calp_\zeta^{\rm iso}$ 
is the isotropic part of the spectrum,
\mbox{$p\equiv(\calp_\|-\calp_+)/\calp_+$} with $\calp_\|$ being the power
spectrum of the perturbations of the longitudinal component of the vector field
and $\calp_+$ being the even part of the power spectra of the perturbations of
the transverse components of the vector field defined as 
\mbox{$\calp_+\equiv\frac12(\calp_{\sf L}+\calp_{\sf R})$}, with {\sf L} and 
{\sf R} denoting the left and right transverse polarisations respectively 
\cite{stanis}. Also, \mbox{$N_A^2\equiv\sum_iN_A^i N_A^i$}, where 
\mbox{$N_A^i\equiv\frac{\partial N}{\partial A_i}$} with $N$ denoting the 
elapsing e-folds and $i$ labelling the spatial components of the vector field
$A_\mu$.

For the vector curvaton mechanism we have \mbox{$N_A=\frac23\hat\Omega_A/A$} 
\cite{stanis}, where \mbox{$\hat\Omega_A\equiv\frac{3\Omega_A}{4-\Omega_A}$}, 
with \mbox{$\Omega_A\equiv\rho_A/\rho$} being the density parameter of the 
vector field when it decays. Combining the above, it is straightforward to show
that
\begin{equation}
\sqrt{\calp_\zeta}
=\frac23\sqrt{\frac{p}{g}}\;\hat\Omega_A
\frac{\sqrt{\calp_+}}{A}
\end{equation}
where $A$ is the modulus of the vector field and we assumed that 
\mbox{$\calp_\zeta^{\rm iso}\simeq\calp_\zeta
$},
because the observations suggest that 
$\calp_\zeta$ is approximately isotropic with 
\mbox{$g\lsim 0.3$}~\cite{GE}, where \mbox{$
\sqrt{\calp_\zeta}=4.8\times 10^{-5}$} is the 
observed curvature perturbation. Now, if particle production is strongly 
anisotropic with \mbox{$\calp_\|\gg\calp_+$} 
\mbox{\{\mbox{$\calp_\|\ll\calp_+$}\}} 
then we have \mbox{$p\simeq\calp_\|/\calp_+$}
\mbox{\{\mbox{$p\simeq -1$}\}} and \mbox{$\delta A\sim\sqrt{\calp_\|}$}
\mbox{\{\mbox{$\delta A\sim\sqrt{\calp_+}$}\}}. Thus, 
\mbox{$
\sqrt{\calp_\zeta}\simeq
\frac23\frac{\hat\Omega_A}{\sqrt{|g|}}\frac{\delta A}{A}$}. 
Alternatively, if particle production is almost isotropic then 
\mbox{$p,g\approx 0$}. In this case, the vector field alone can be responsible 
for the curvature perturbation. Then, 
\mbox{$\sqrt{\calp_\zeta}\approx
\zeta=\hat\Omega_A\zeta_A$}, with 
\mbox{$\zeta_A=\frac23\frac{\delta A}{A}$} being the curvature perturbation 
attributed to the vector curvaton \cite{stanis}. Thus, in all cases, we can 
write
\begin{equation}
\sqrt{\calp_\zeta}\simeq
\frac23C\hat\Omega_A\frac{\delta A}{A},
\label{zC}
\end{equation}
where \mbox{$C=1/\sqrt{|g|}$} \{\mbox{$C=1$}\} for strongly anisotropic
\{approximately isotropic\} particle production of the vector bosons during
inflation.

In our case, the role of $A_\mu$ above is played by the supermassive vector 
boson $X_\mu$. Then, 
using Eq.~(\ref{Xew})
, we obtain
\begin{equation}
\delta X_{\rm end}\sim
\frac{
\sqrt{\calp_\zeta}}{C\sqrt{\Omega_X}}
\frac{m_PH_*}{m_X}\,,
\label{dYew}
\end{equation}
where we used that \mbox{$\hat\Omega_X\simeq\Omega_X$}.

Let us investigate now the corresponding primordial magnetic field (PMF) which
results from the projection of $B^Y$ onto the photon direction. Similarly to
$B^Y$, we have \mbox{{\boldmath $B$}$\,\equiv\,$%
{\boldmath $\nabla$}$\,\times\,\delta${\boldmath $A$}}. Hence, at galaxy 
formation (denoted by `gf'), the rms PMF over the lengthscale $\ell$ is
\begin{equation}
B_{\rm seed}(\ell)\sim 10^2\,\frac{\delta A_{\rm gf}}{\ell}\,,
\label{Bseed}
\end{equation}
where we considered that, because of gravitational collapse, flux conservation
amplifies the PMF by a factor 
\mbox{($\frac{0.1\;{\rm Mpc}}{10\;{\rm kpc}}$)$^2\sim 10^2$}, 
with 0.1~Mpc \{10~kpc\} being the typical dimensions of a protogalaxy before
\{after\} the collapse at the time of galaxy formation. 
As the Universe expands, flux conservation reduces the magnitude
of the magnetic field as \mbox{$B\propto a^{-2}$}. Since \mbox{$\ell\propto a$}
before galaxy formation, we find \mbox{$\delta A\propto a^{-1}$}. 
Scaling back to the end of inflation and, assuming (for simplicity) prompt 
reheating
, we obtain
\begin{equation}
\!\,
B_{\rm seed}(\ell)\sim 
10^2\frac{\cos\theta_W}{(1+z_{\rm gf})^{-1}}
\frac{T_{_{\rm CMB}}}{T_{\rm end}}
\frac{\delta Y_{\rm end}}{\ell}\,,\hspace{-1cm}
\label{Bseed+}
\end{equation}
where 
\mbox{$T_{_{\rm CMB}}\sim 0.23\,$meV} is the temperature of the CMB and 
$T_{\rm end}=(\frac{30}{\pi^2g_*})^{1/4}M
\sim 10^{15}\,$GeV 
is the temperature at the end of inflation, with \mbox{$z_{\rm gf}\simeq 10$} 
being the redshift at galaxy formation. 

Both the perturbations of the supermassive $X$-boson and the hypercharge field
at the end of inflation are due to the perturbation of the unbroken GUT bosons
$\delta W$. Hence, we expect 
\mbox{$\delta X_{\rm end}\sim\delta W_{\rm end}\sim\delta Y_{\rm end}$}.%
\footnote{For example, if we take as our GUT Flipped SU(5): SU(5)$\times$U(1) 
as in Ref.~\cite{mine}, we have for the supermassive $V$-boson
\mbox{$V_\mu\propto W_\mu^{(15)}-\tan\Theta\,W_\mu^{(0)}$} and for the 
hyperhcarge \mbox{$Y_\mu\propto W_\mu^{(15)}+\cot\Theta\,W_\mu^{(0)}$}, where
$W_\mu^{(15)}$ and $W_\mu^{(0)}$ belong to the SU(5) and the U(1) constituent 
groups of the GUT respectively and \mbox{$\tan\Theta\equiv \bar g/g$} with 
$\bar g$ and $g$ being the gauge couplings of the SU(5) and the U(1) parts of 
the GUT. Therefore, if particle production generates a superhorizon spectrum 
for the Abelian vector boson $W_\mu^{(0)}$ then 
\mbox{$-\cot\Theta\,\delta V_\mu=\delta W_\mu^{(0)}=\tan\Theta\,\delta Y_\mu$}
\cite{mine}.}
Using this and combining Eqs.~(\ref{dYew}) and (\ref{Bseed+}) we find
\begin{equation}
\!\,
B_{\rm seed}(\ell)\sim 10^3
\frac{T_{_{\rm CMB}}}{T_{\rm end}}
\frac{
\sqrt{\calp_\zeta}}{C\sqrt{\Omega_X}}
\frac{m_PH_*}{m_X\ell}\,.\hspace{-1cm}
\label{seed}
\end{equation}
Putting the numbers in we arrive at
\begin{equation}
B_{\rm seed}(\ell)\sim 
\frac{10^{-29}}{C\sqrt{\Omega_X}}
\left(\frac{1\;{\rm kpc}}{\ell}\right){\rm G}\,.
\label{B}
\end{equation}
Thus, we see that, at the scale of the largest turbulent eddy 
\mbox{$\ell\gsim 100\,$pc}, we can obtain 
\mbox{$B_{\rm seed}\sim 10^{-28}\,$G}
if \mbox{$C\sqrt{\Omega_X}\sim 1$}. 
%
In particular, if particle production is (approximately) isotropic, 
\mbox{$C=1$}. The vector curvaton alone can produce $\zeta$ and we have 
the bound \mbox{$\Omega_X\gsim 10^{-2}$} in order to avoid excessive 
non-Gaussianity \cite{varkin}. Thus, we obtain 
\mbox{$B_{\rm seed}(100\,{\rm pc})\lsim 10^{-27}\,$G}.
Similarly, if particle production is strongly
anisotropic, \mbox{$C=1/\sqrt{|g|}$}. Demanding that \mbox{$\delta X/X\lsim 1$}
in order for the perturbative approach to be valid, Eq.~(\ref{zC}) suggests
\mbox{$\Omega_X\gsim\sqrt{|g|\,\calp_\zeta}
\gsim 10^{-5}$}, using that 
\mbox{$|g|\gsim 0.02$} in order for statistical anisotropy to be observable 
\cite{planck}. Therefore, Eq.~(\ref{B}) gives again
\mbox{$B_{\rm seed}(100\,{\rm pc})\lsim 10^{-27}\,$G}.
Note that, when particle
production is strongly anisotropic we expect \mbox{$\Omega_X\ll 1$} because,
otherwise, statistical anisotropy would be too large \cite{stanis} 
(see also Ref.~\cite{jack}).

The above PMF is just about strong and coherent enough to seed the galactic 
dynamo mechanism, which explains the observed $\mu$G magnetic fields of the
galaxies \cite{acd}.\footnote{The galactic magnetic fields can also magnetise
the intergalactic medium by expulsion, e.g. through the Parker 
instability~\cite{parker}.}
Note that, being due to the same cause (the perturbations
of the $W$ bosons of the unbroken GUT), the PMF and the curvature perturbation 
(in the isotropic case) are correlated, which implies that overdensities are 
more intensely magnetised than the surrounding matter. This may assist 
gravitational collapse and the formation of Population~III stars and quasars by
removing angular momentum \cite{ang}.


A number of mechanisms exist in the literature for the generation the flat 
superhorizon spectrum of perturbations of the unbroken GUT gauge bosons, which 
we postulated at the beginning. 
In Ref.~\cite{vecurv}
it is shown that an Abelian vector field obtains the desired 
perturbation spectrum if its effective mass during inflation is
\mbox{$m^2\approx -2H_*^2$}. Such an effective mass-squared can be generated by
coupling the vector field non-minimally to gravity with an $\frac16RA^2$ 
term \cite{TW}. This possibility is explored in Refs.~\cite{stanis} and 
\cite{nonmin}, where it was shown that \mbox{$\calp_\|=2\calp_+$}, i.e. 
\mbox{$p=1$}. Such a vector curvaton could only generate statistical anisotropy
in $\zeta$.
A different idea for the generation of a flat superhorizon spectrum of 
curvature perturbations is studied in Refs.~\cite{varkin,sugravec}, this 
time based on a non-trivial evolution for the gauge kinetic function 
\mbox{$f\propto a^{-1\pm 3}$} during inflation, which is natural in 
supergravity theories and which is an attractor solution if $f$ is modulated by
the inflaton field \cite{attr}.\footnote{PMF generation with this mechanism is 
considered in Ref.~\cite{gaugekin}. The same kind of model has 
been employed in Ref.~\cite{anisinf} to produce weakly anisotropic inflation 
which can generate statistical anisotropy in the perturbations of the inflaton 
field.} If the vector curvaton is massless during inflation then 
\mbox{$\calp_\|\rightarrow 0$} and \mbox{$p=-1$}. However, one can consider a 
massive vector curvaton. Then, in Ref.~\cite{varkin}, it was shown that scale 
invariance is obtained when \mbox{$m\propto a$}. In this case, if the vector 
curvaton remains light, \mbox{$p\simeq\calp_\|/\calp_+\gg 1$} and it can only 
generate statistical anisotropy in $\zeta$. However, if the vector curvaton 
becomes heavy by the end of inflation then particle production is rendered 
isotropic with \mbox{$p\approx 0$}, so that the vector curvaton alone can 
generate $\zeta$.
For a recent review of vector curvaton models see Ref.~\cite{VCrev}.\footnote{%
One other possibility to generate a superhorizon spectrum of perturbations
for a vector boson field is by employing the axial coupling 
$\alpha F_{\mu\nu}\tilde F^{\mu\nu}$ as explored in Ref.~\cite{axial}.
Application of this possibility in the vector curvaton mechanism is studied in 
Ref.~\cite{axvecurv}.}.

In passing we note that the effect of the presented mechanism can be indirectly
witnessed by observational evidence on structure formation, because, as 
mentioned, overdensities may be more magnetised, which assists graviational 
collapse (by removal of angular momentum). However, on the magnetic field side,
any direct evidence is inhibited by the sweeping action of the galactic dynamo.
Magnetised overdensities that do not undergo dynamo amplification are too 
weakly magnetised to offer any clear observational test.

In summary, in the context of GUT hybrid inflation, we have explored a 
particular vector curvaton scenario which employs the GUT gauge bosons to 
affect (by generating statistical anisotropy) or even to produce the observed 
curvature perturbation $\zeta$, while simultaneously generating a primordial 
magnetic field strong and coherent enough to account for galactic magnetism. 
Assuming that inflation gives rise to a flat superhorizon spectrum of 
perturbations for (at least some of) the GUT bosons, we have shown that they
can affect or generate $\zeta$ at the breaking of grand unification (which also
terminates inflation), when a supermassive GUT boson can act as a vector 
curvaton field. The spectrum of the corresponding hypercharge perturbations 
survives the hot big bang in the form of a hypermagnetic field, which is frozen
into the highly conducting plasma. The projection of the hypercharge onto the 
photon direction at the electroweak phase transition produces a primordial 
magnetic field, which can be strong and coherent enough to seed the galactic 
dynamo mechanism and explain the observed magnetic fields of the galaxies. The 
generation of the original flat superhorizon spectrum of perturbations for the 
gauge bosons of the unbroken GUT during inflation is due to some other 
mechanism which we simply postulated. However, concrete such mechanisms exist 
in the~literature. 

\section*{acknowledgements}
I wish to thank the University of Crete for the hospitality. This work was
supported (in part) by the Lancaster-Manchester-Sheffield Consortium for 
Fundamental Physics under STFC grant ST/J000418/1.

\begin{thebiblio
}{99}

\bibitem{book}
A.~R.~Liddle and D.~H.~Lyth,
{\em The Primordial Density Perturbation: 
Cosmology, Inflation and the origin of Structure},
(Cambridge University Press, Cambridge U.K., 2009).

\bibitem{hawking}
  G.~W.~Gibbons and S.~W.~Hawking,
  Phys.\ Rev.\  D {\bf 15} (1977) 2738.

\bibitem{vecurv}
K.~Dimopoulos,
  Phys.\ Rev.\  D {\bf 74} (2006) 083502.

\bibitem{yokosoda}
S.~Yokoyama and J.~Soda,
  JCAP {\bf 0808} (2008) 005.

\bibitem{stanis}
K.~Dimopoulos, M.~Karciauskas, D.~H.~Lyth and Y.~Rodriguez,
  JCAP {\bf 0905} (2009) 013.

\bibitem{fnlanis}
M.~Karciauskas, K.~Dimopoulos and D.~H.~Lyth,
  Phys.\ Rev.\  D {\bf 80} (2009) 023509.

\bibitem{anisinf}
C.~Pitrou, T.~S.~Pereira and J.~P.~Uzan,
  JCAP {\bf 0804} (2008) 004;
M.~a.~Watanabe, S.~Kanno and J.~Soda,
  Phys.\ Rev.\ Lett.\  {\bf 102} (2009) 191302;
T.~R.~Dulaney and M.~I.~Gresham,
  Phys.\ Rev.\  D {\bf 81} (2010) 103532;
A.~E.~Gumrukcuoglu, B.~Himmetoglu and M.~Peloso,
  Phys.\ Rev.\  D {\bf 81} (2010) 063528.

\bibitem{varkin}
K.~Dimopoulos, M.~Karciauskas and J.~M.~Wagstaff,
  Phys.\ Rev.\  D {\bf 81} (2010) 023522;
  Phys.\ Lett.\  B {\bf 683} (2010) 298.

\bibitem{curv}
S.~Mollerach,
  Phys.\ Rev.\  D {\bf 42} (1990) 313;
A.~D.~Linde and V.~F.~Mukhanov,
  Phys.\ Rev.\  D {\bf 56} (1997) 535;
D.~H.~Lyth and D.~Wands,
Phys.\ Lett.\ B {\bf 524} (2002) 5.

\bibitem{hybrid}
A.~D.~Linde,
  Phys.\ Rev.\  D {\bf 49} (1994) 748;
%
E.~J.~Copeland, A.~R.~Liddle, D.~H.~Lyth, E.~D.~Stewart and D.~Wands,
  Phys.\ Rev.\  D {\bf 49} (1994) 6410;
G.~R.~Dvali, Q.~Shafi and R.~K.~Schaefer,
  Phys.\ Rev.\ Lett.\  {\bf 73} (1994) 1886;
G.~Lazarides, R.~K.~Schaefer and Q.~Shafi,
  Phys.\ Rev.\  D {\bf 56} (1997) 1324;
C.~Panagiotakopoulos,
  Phys.\ Rev.\  D {\bf 55} (1997) 7335.

\bibitem{kron}
P.~P.~Kronberg,
  Rept.\ Prog.\ Phys.\  {\bf 57} (1994) 325.

\bibitem{beck}
R.~Beck, A.~Brandenburg, D.~Moss, A.~Shukurov and D.~Sokoloff,
  Ann.\ Rev.\ Astron.\ Astrophys.\  {\bf 34} (1996) 155;
L.~M.~Widrow,
  Rev.\ Mod.\ Phys.\  {\bf 74} (2003) 775.

\bibitem{pmfrev}
D.~Grasso and H.~R.~Rubinstein,
  Phys.\ Rept.\  {\bf 348} (2001) 163;
M.~Giovannini,
  Int.\ J.\ Mod.\ Phys.\  D {\bf 13} (2004) 391.

\bibitem{TW}
M.~S.~Turner and L.~M.~Widrow,
  Phys.\ Rev.\  D {\bf 37} (1988) 2743.

\bibitem{mine}
A.~C.~Davis, K.~Dimopoulos, T.~Prokopec and O.~Tornkvist,
  Phys.\ Lett.\  B {\bf 501} (2001) 165;
K.~Dimopoulos, T.~Prokopec, O.~Tornkvist and A.~C.~Davis,
  Phys.\ Rev.\  D {\bf 65} (2002) 063505.

\bibitem{pmfinf}
W.~D.~Garretson, G.~B.~Field and S.~M.~Carroll,
  Phys.\ Rev.\  D {\bf 46} (1992) 5346;
F.~D.~Mazzitelli and F.~M.~Spedalieri,
  Phys.\ Rev.\  D {\bf 52} (1995) 6694;
A.~Dolgov,
  Phys.\ Rev.\  D {\bf 48} (1993) 2499;
B.~Ratra,
  Astrophys.\ J.\  {\bf 391} (1992) L1;
E.~A.~Calzetta, A.~Kandus and F.~D.~Mazzitelli,
  Phys.\ Rev.\  D {\bf 57} (1998) 7139;
O.~Bertolami and D.~F.~Mota,
  Phys.\ Lett.\  B {\bf 455} (1999) 96;
M.~Giovannini,
  Phys.\ Rev.\  D {\bf 62} (2000) 123505;
T.~Prokopec and E.~Puchwein,
  JCAP {\bf 0404} (2004) 007;
  Phys.\ Rev.\  D {\bf 70} (2004) 043004;
K.~Enqvist, A.~Jokinen and A.~Mazumdar,
  JCAP {\bf 0411} (2004) 001;
M.~R.~Garousi, M.~Sami and S.~Tsujikawa,
  Phys.\ Lett.\  B {\bf 606} (2005) 1;
A.~Ashoorioon and R.~B.~Mann,
  Phys.\ Rev.\  D {\bf 71} (2005) 103509;
J.~E.~Madriz Aguilar and M.~Bellini,
  Phys.\ Lett.\  B {\bf 642} (2006) 302;
L.~Campanelli, P.~Cea, G.~L.~Fogli and L.~Tedesco,
  Phys.\ Rev.\  D {\bf 77} (2008) 043001;
  Phys.\ Rev.\  D {\bf 77} (2008) 123002.

\bibitem{gaugekin}
M.~Giovannini,
  Phys.\ Rev.\  D {\bf 64} (2001) 061301;
K.~Bamba and J.~Yokoyama,
  Phys.\ Rev.\  D {\bf 69} (2004) 043507;
  Phys.\ Rev.\  D {\bf 70} (2004) 083508;
O.~Bertolami and R.~Monteiro,
  Phys.\ Rev.\  D {\bf 71} (2005) 123525;
J.~M.~Salim, N.~Souza, S.~E.~Perez Bergliaffa and T.~Prokopec,
  JCAP {\bf 0704} (2007) 011;
K.~Bamba and M.~Sasaki,
  JCAP {\bf 0702} (2007) 030;
J.~Martin and J.~Yokoyama,
  JCAP {\bf 0801} (2008) 025;
K.~Bamba and S.~D.~Odintsov,
  JCAP {\bf 0804} (2008) 024;
K.~Bamba, C.~Q.~Geng and S.~H.~Ho,
  JCAP {\bf 0811} (2008) 013;
S.~Kanno, J.~Soda and M.~a.~Watanabe,
  JCAP {\bf 0912} (2009) 009.


\bibitem{mukh}
V.~Demozzi, V.~Mukhanov and H.~Rubinstein,
  JCAP {\bf 0908} (2009) 025.

\bibitem{attr}
J.~M.~Wagstaff and K.~Dimopoulos, 
  Phys.\ Rev.\  D {\bf 83} (2011) 023523.

\bibitem{jack}
K.~Dimopoulos, G.~Lazarides and J.~M.~Wagstaff,
  JCAP {\bf 1202} (2012) 018.

\bibitem{sodarev}
J.~Soda,
  Class.\ Quant.\ Grav.\  {\bf 29} (2012) 083001.

\bibitem{KT}
E.~W.Kolb and M.~S.~Turner,
{\em The Early Universe},
(Frontiers in Physics; v.~69, Addison-Wesley Publishing Company, 
Reading Massachusetts U.S.A., 1994).

\bibitem{nonAbelmass}
A.~Billoire, G.~Lazarides and Q.~Shafi,
  Phys.\ Lett.\  B {\bf 103} (1981) 450;
T.~A.~DeGrand and D.~Toussaint,
  Phys.\ Rev.\  D {\bf 25} (1982) 526.

\bibitem{GE}
N.~E.~Groeneboom and H.~K.~Eriksen,
  Astrophys.\ J.\  {\bf 690} (2009) 1807;
N.~E.~Groeneboom, L.~Ackerman, I.~K.~Wehus and H.~K.~Eriksen,
  Astrophys.\ J.\  {\bf 722} (2010) 452;
D.~Hanson and A.~Lewis,
  Phys.\ Rev.\  D {\bf 80} (2009) 063004.

\bibitem{planck}
A.~R.~Pullen and M.~Kamionkowski,
  Phys.\ Rev.\  D {\bf 76} (2007) 103529.

\bibitem{acd}
A.~C.~Davis, M.~Lilley and O.~Tornkvist,
  Phys.\ Rev.\  D {\bf 60} (1999) 021301.

\bibitem{parker}
 E.~N.~Parker,
  Astrophys.\ J.\  {\bf 122} (1955) 293;
  Astrophys.\ J.\  {\bf 154} (1968) 49;
  Space Sci. Rev. {\bf 9} (1969) 651;
  Astrophys.\ J.\  {\bf 163} (1971) 255.

\bibitem{ang}
C.~J.~Hogan, Phys.\ Rev.\ Lett.\  {\bf 51} (1983) 1488; 
L.~Mestel, Physica Scripta {\bf T11} (1985) 53.

\bibitem{nonmin}
K.~Dimopoulos and M.~Karciauskas,
  JHEP {\bf 0807} (2008) 119.

\bibitem{sugravec}
K.~Dimopoulos,
  Phys.\ Rev.\  D {\bf 76} (2007) 063506.

\bibitem{VCrev}
K.~Dimopoulos,
1107.2779 [hep-ph], 
Int. J. of Mod. Phys. D {\bf 21} (2012) 1250023.

\bibitem{axial}
M.~M.~Anber and L.~Sorbo,
  JCAP {\bf 0610} (2006) 018;
L.~Sorbo,
  JCAP {\bf 1106} (2011) 003;
R.~Durrer, L.~Hollenstein and R.~K.~Jain,
  JCAP {\bf 1103} (2011) 037.

\bibitem{axvecurv}
K.~Dimopoulos and M.~Karciauskas,
  JHEP {\bf 1206} (2012) 040.

\end{thebiblio}


\end{document}